\documentclass[prl,twocolumn,floats,aps,epsfig,nofootinbib,amssymb]{revtex4}
\usepackage{graphicx}
\usepackage{cancel}
\usepackage{amssymb}
\usepackage{textcomp}
\usepackage{amsmath}
\usepackage{bm}
\usepackage{times}
\usepackage{epsfig}
\usepackage{color}
\usepackage{graphics}
\usepackage{hyperref}
\newcommand{\beq} {\begin{equation}}
\newcommand{\eeq} {\end{equation}}
\newcommand{\ba} {\begin{eqnarray}}
\newcommand{\ea} {\end{eqnarray}}
\usepackage{setspace}

\begin{document}
\newcount\hour \newcount\minute
\hour=\time \divide \hour by 60
\minute=\time
\count99=\hour \multiply \count99 by -60 \advance \minute by \count99
\newcommand{\mydate}{\ \today \ - \number\hour :00}
%
\newcommand\plot[2]{\includegraphics[width=#1, angle=0]{plots/#2}}
\newcommand\Fdiag[2]{\includegraphics[#1]{diags/#2}}
%
\title{\Large{Gauge Origin of $M$-Parity and the $\mu$-Term in Supersymmetry}}
\author{Pavel Fileviez P\'erez} 
\affiliation{Phenomenology Institute (PHENO), Department of Physics, University of Wisconsin-Madison, WI 53706, USA \\
Center for Cosmology and Particle Physics (CCPP), Department of Physics, New York University, NY, 10003, USA}
\author{Mart\'in Gonz\'alez-Alonso}
\affiliation{Theoretical Nuclear, Particle, Astrophysics, and Cosmology (NPAC) Group, Department of Physics, University of Wisconsin-Madison, WI 53706, USA}
\author{Sogee Spinner}
\affiliation{Phenomenology Institute (PHENO), Department of Physics, University of Wisconsin-Madison, WI 53706, USA}
\date{\today}
\begin{abstract}
In this article we present a simple theoretical framework where the origin of the $\mu$-term and the matter-parity violating interactions of the minimal supersymmetric standard model 
can be understood from the spontaneous breaking of new Abelian gauge symmetries. In this context the masses of the $Z'$ gauge bosons, the $M$-parity violating scale and the $\mu$-term are determined by the 
supersymmetry breaking scale. The full spectrum of the theory is discussed in detail. We investigate the predictions for the Higgs masses in detail showing that it is possible to satisfy the LEP2 bounds 
even with sub-TeV squark masses. The model predicts the existence of light colored fields, lepton and baryon number violation, and new neutral gauge bosons at the Large Hadron Collider.   
\end{abstract}
\maketitle
\begin{spacing}{1.15}
\tableofcontents
\end{spacing}
\section{I. Introduction}
Whether low energy supersymmetry (SUSY) in the guise of the minimal supersymmetric standard model (MSSM) is a good description of nature or not is currently being aggressively tested at the Large Hadron Collider (LHC).  
One thing that is certain is that it has withstood the test of time as a strong candidate for new physics in the minds of many due to its elegant solution to the hierarchy problem, the unification of the gauge couplings and its accommodation of a dark matter candidate. The current proliferation of experimental results make this an exciting time to consider non-canonical SUSY phenomenologies, especially when such phenomenologies stem from solutions to fundamental issues in SUSY. For a review of the phenomenological aspects of the MSSM see Ref.~\cite{Drees}. 

In this paper we will address two such issues in the context of a single model and outline some of its interesting phenomenology. The first is the status of the baryon- and lepton-number violating terms allowed by the gauge symmetries of the MSSM but whose presence, in general, would lead to unacceptably fast proton decay. The second is the so-called $\mu$-problem, referring to the only dimensionful parameter in the MSSM superpotential (the mass term for the Higgsinos), whose value can be expected to be arbitrarily large but must be fixed at or below the SUSY scale for successful electroweak symmetry breaking.

Typically, one appeals to discrete symmetries to fix both issues. Proton decay is typically assumed to be absent due to the discrete $M$-parity (or $R$-parity) symmetry which forbids tree-level baryon and lepton number violating terms while also guaranteeing the stability of the lightest supersymmetric particle (LSP). This has important consequences for both colliders (detectable missing energy) and cosmology (the LSP is a candidate for the dark matter of the universe). Meanwhile, a discrete $Z_3$ symmetry is typically imposed to forbid the bilinear $\mu$-term, which is replaced by singlet field, whose vacuum expectation value (VEV) generates the $\mu$-term after symmetry breaking. This model is referred to as the next to minimal supersymmetric standard model (NMSSM) and it and its deviations are reviewed in Ref.~\cite{NMSSMreview}.  Such a scenario expands the Higgs sector thereby potentially changing expectation for collider physics and causing cosmological concerns related to domain walls.

Our approach in this paper is to understand the possible origin of the discrete symmetries mentioned above from the spontaneous breaking of local symmetries. 
While it is maybe true that this simply amounts to replacing one symmetry by another, we think that the corresponding $Z'$ gauge bosons associated with local symmetries 
allow for a better handle on testing such ideas. Therefore, we propose a simple model where the origin of the $\mu$-term and the matter-parity violating interactions of the MSSM can be understood from the spontaneous breaking of two new Abelian gauge symmetries: $U(1)_{B-L}$ and $U(1)_S$ where only the third generation carries $U(1)_S$ charge. In order to define an anomaly free theory new colored triplets exotics are needed. $B-L$ is broken by the VEV of the ``right-handed" sneutrino giving rise to lepton number violating $M$-parity violation and 
$U(1)_S$ is broken by the VEV of a SM singlet, $S$, which generates the $\mu$-term. The new $Z'$ associated with $U(1)_S$ give rise to flavor violation without experimental conflict. 
Symmetry breaking also allows for a consistent scenario for fermion masses, predicting a very small mixings between the third generation and the others. The numerical 
predictions for the lightest Higgs boson are investigated up to one-loop level showing the possibility to satisfy the experimental bounds from the LEP2 experiment with squark masses below 1 TeV. 
Finally, we make a brief discussion of how one could observe lepton and baryon number violation at the LHC in agreement with the experimental bounds on proton decay.

The remainder of this article is organized as follows: In Section II we expand on the issues of $M$-parity and the $\mu$-term and past attempts to address them. 
In Section III we propose our new theoretical framework where both issues can be solved and discuss the necessary symmetry breaking in Section IV. 
The properties of the full spectrum are presented in Section V, while in Section VI the main phenomenological aspects are presented. Our findings are summarizes in Section VII.
\section{II. The $\mu$-problem and $M$-parity}
As mentioned above, different approaches to the $\mu$-problem and $M$-parity have significantly different consequences and it's especially the presence or absence 
of the latter that answers one of the most important questions of the MSSM: the stability of the LSP.  A brief review is therefore in order.
\subsection{A. $M$-Parity Violating Interactions}
The fate of M-parity in the MSSM has important cosmological and phenomenological implications. 
M-parity is defined as $M=(-1)^{3(B-L)}$, where $B$ and $L$  stand for total baryon number and lepton number, respectively.  In general, the MSSM contains lepton and baryon number violating interactions in the superpotential:
\begin{eqnarray} 
{\cal W}_{MV} &=& \epsilon \hat{L} \hat{H}_u \ + \ \lambda \hat{L} \hat{L} \hat{e}^c 
\ + \  \lambda^{'} \hat{Q} \hat{L} \hat{d}^c \ + \  \lambda^{''} \hat{u}^c \hat{d}^c \hat{d}^c
\end{eqnarray}
In most phenomenological studies it is assumed that $M$-parity is conserved by hand, \textit{i.e.}  the above interactions are absent, or that only some of them are present: explicit $M$-parity breaking. 
Since these terms affect the most significant features of the MSSM, the origin of $M$-parity conservation or violation must be understood dynamically. It has long been realized that the simplest forum for this is $B-L$  symmetric theories~\cite{RpC}. Since $M$-parity is a subgroup of $B-L$, at the $B-L$ scale all the above interactions are absent. Local $B-L$ further requires the existence of right-handed neutrinos for anomaly cancellation which also provide the most minimal way of breaking $B-L$~\cite{PRL}: the VEV of the right-handed sneutrino\footnote{A scenario further motivated by string theory~\cite{Braun:2005nv, Ambroso:2009jd}}. Therefore, in the simplest theory of $M$-parity, it is spontaneously broken and the B-L and the M-parity violating scales are determinate by the soft SUSY breaking scale. As has been emphasized in Ref.~\cite{PRL} after symmetry breaking only bilinear lepton number violating interactions are present and there are no dimension four contributions to proton decay.  
For a review on proton decay  see Ref.~\cite{Reviewpdecay}. 
\subsection{B. The $\mu$-Problem and New Symmetries} 
The $\mu$ parameter is part of the $M$-parity conserving MSSM superpotential:
\begin{eqnarray}
{\cal W}_{MC} &=& Y_u \hat{Q} \hat{H}_u \hat{u}^c \ + \ Y_d \hat{Q} \hat{H}_d \hat{d}^c \nonumber \\
& + & Y_e \hat{L} \hat{H}_d \hat{e}^c \ + \ \mu \hat{H}_u \hat{H}_d,
\end{eqnarray}
and defines the mass of the Higgsinos and plays a very important role in electroweak symmetry breaking.This relates the $Z$ boson mass (which we can use to define the weak scale), the $\mu$ term and the soft terms in the Higgs sector:
\begin{equation}
\label{EWSB}
	\frac{1}{2} M_{Z}^2 = - |\mu|^2 \ - \  \left( \frac{m_{H_u}^2 \tan^2 \beta - m_{H_d}^2}{ \tan^2 \beta - 1 } \right),
\end{equation} 
where $m_{H_u}$ and $m_{H_d}$ are the soft terms for the MSSM Higgses and $\tan \beta = v_u/v_d$. Notice that in order to satisfy the above equation the second term on the right-hand side must be negative and its magnitude must be larger than the $\mu$-term, for large $\tan \beta$, this translates into the condition that $\mu$ must be smaller in magnitude than the soft terms. At the same time $\mu$ is a mass dimensionful parameter in the superpotential and in principle it could be very large. This is the so-called $\mu$-problem. From chargino searches the $\mu$ lower bound is approximately $\mu \gtrsim 100$ GeV. 

Many scenarios have been proposed to explain the origin of a SUSY-scale $\mu$-term~\cite{muterm1,muterm2,muterm3,Aoki,muterm4, Cheng:1998nb}. In the NMSSM one introduces a new singlet, $S$, and replaces the $\mu$ term in the superpotential 
by the term $\lambda \hat{S} \hat{H}_u \hat{H}_d$. Then, the $\mu$-parameter is defined by the VEV of $S$ which is around the SUSY scale. In order to achieve this scenario a new discrete symmetry, a $Z_3$ symmetry, 
is introduced which forbids the mass term in the superpotential. See Ref.~\cite{NMSSMreview} for a review of the NMSSM.  However, the question of a dynamical origin for the $\mu$-term remains.

As in the $M$-parity case it is possible to find a gauge origin to the $\mu$-term by introducing a new abelian symmetry which is spontaneously broken at the TeV scale.  However, unlike $B-L$ for $M$-parity, it is hard to pinpoint the simplest model.  Various possibilities have been investigated by many groups~\cite{muterm3, muterm4}.  Since the $Z_3$ symmetry is replaced by a gauge symmetry, the cosmological problems associated with the spontaneous breaking of the discrete symmetry is avoided.  We see such an approach as appealing because it connects the $\mu$ term to the existence of a new gauge boson which could experimentally relate to the mechanism for the dynamical generation of the $\mu$-term.

Combining the two possible solutions, discrete and local symmetries, to these two issues of $M$-parity and the $\mu$-problem affords four different frameworks for approaching these issues.  Typically, most of the phenomenological studies have been performed in a model 
where a $Z_2$ (matter parity) and $Z_3$ is assumed. A second possibility is a simple extension of the model in Ref.~\cite{PRL}, where $M$-parity is spontaneously broken along with $B-L$ and a $Z_3$ symmetry is assumed 
to explain the $\mu$-term. A third scenario was quoted as an example in Ref.~\cite{muterm3} where the generation of the $\mu$-term is defined by the scale where a new $U(1)^{'}$ symmetry 
is broken and a $Z_2$ symmetry is assumed to avoid fast proton decay. Finally, one can consider a more complete framework with two Abelian symmetries for understanding dynamically the generation of the $M$-parity violating terms 
and the $\mu$-term.

The difficulty in flagging a simplest gauge solution to the $\mu$-problem is due to three issues that usually arise.  Since the Higgs fields will now have a new charge, it is not \textit{a priori} clear that Yukawa couplings generating fermion masses will be gauge invariant thus making fermion mass generation non-trivial.  Anomaly cancellation usually requires the existence of new exotic color states.  These will either couple to matter and induce rapid proton decay or form a separate sector with no couplings to matter resulting in the lightest exotic being stable.  The latter scenario would lead to relic bound states, which could disagree with current cosmological data.  We have found that several papers in the literature contain such traits, with a noteworthy example of the last one being Ref.~\cite{Aoki}, which solves both issues in a nice way but contains stable colored particles.

Due to these possible complications, we take this opportunity to state our goals in addressing the $\mu$-term:
\begin{itemize}
	\item No dimensionful parameters in the superpotential which would affect the electroweak symmetry breaking (EWSB) condition, Eq.~(\ref{EWSB}).  This includes the $\mu$-term as well as the $\epsilon$-term.
	\item Explain the long lifetime of the proton.
	\item No stable colored fields.
	\item Generation of all fermion masses and mixings.
\end{itemize}

Now, we are ready to discuss the simplest theoretical framework where these issues are addressed.
\section{III. Theoretical Framework}
In order to investigate how $M$-parity violating terms and the $\mu$-term are generating dynamically we introduce two extra Abelian symmetries, $U(1)_{B-L}$ and $U(1)_S$. The first is 
needed to understand the origin of $M$-parity while the second symmetry governs how the $\mu$-term is generated. Therefore, the model will be based on the local gauge symmetry
\begin{equation}
	SU(3)_C \otimes SU(2)_L \otimes U(1)_Y \otimes U(1)_{B-L} \otimes U(1)_S
\end{equation}
Inspired by the $\mathbf{27}$ of $E_6$, we introduce four new fields: three generations of right-handed neutrinos necessary to gauge $B-L$, $S$ whose VEV generates the $\mu$-term and $T$ and $\bar T$ needed to cancel the $U(1)_S$ anomalies.  

We assume a non-zero $z$ charge (the charge under $U(1)_S$) 
only for the third generation so that only one set of the latter three fields need be introduced (as opposed to one per generation).   This further restricts the coupling of the exotics to third generation fermions only, significantly suppressing their contribution to proton decay. The anomaly cancellation conditions can be satisfied by the charges in Table~\ref{charges}, where $z_u$ is the charge of $\hat{u}^c_3$ and $z_T$ is the charge of $T$. These are the most general charges given the additional 
assumption that the top mass term is gauge invariant.
\begin{center}
\begin{table*}[th!]
\caption{Field Content ($a = 1..2$).}
\begin{center}
\begin{tabular}{|c|c|c|c|c|c|}
\hline
$\ $Field $\ $ & $\ SU(3)\ $ & $ \ SU(2) \ $ & $ \ U(1)_Y \ $ & $ \ U(1)_{B-L} \ $ & $ \quad \quad \quad U(1)_S \quad \quad \quad $
\\
\hline
$\hat{Q}_a$ & 3 & 2 & 1/6 & 1/3 & 0
\\
$\hat{u}^c_a$ & $\bar 3$ & 1 & -2/3 & -1/3 & 0
\\
$\hat{d}^c_a$ & $\bar 3$ & 1 & 1/3 & -1/3 & 0
\\
$\hat{L}_a$ & $1$ & 2 & -1/2 & -1 & 0
\\
$\hat{e}^c_a$ & $1$ & 1 & 1 & 1 & 0
\\
$\hat{\nu}^c_a$ & $1$ & 1 & 0 & 1 & 0
\\
\hline
$\hat{Q}_3$ & 3 & 2 & 1/6 & 1/3 & $\frac{1+z_T}{2}$
\\
$\hat{u}^c_3$ & $\bar 3$ & 1 & -2/3 & -1/3 & $z_u$
\\
$\hat{d}^c_3$ & $\bar 3$ & 1 & 1/3 & -1/3 & $-z_u - z_T$
\\
$\hat{L}_3$ & $1$ & 2 & -1/2 & -1 & $-\frac{1 + 3 Z_T}{2}$
\\
$\hat{e}^c_3$ & $1$ & 1 & 1 & 1 & $1 - z_u + z_T$
\\
$\hat{\nu}^c_3$ & $1$ & 1 & 0 & 1 & $1 + z_u + 2 z_T$
\\
\hline
$\hat{H}_d$ & $1$ & 2 & -1/2 & 0 & $\frac{1}{2} (-1 + 2 z_u + z_T)$
\\
$\hat{H}_u$ & $1$ & 2 & 1/2 & 0 & $\frac{1}{2} (-1 - 2 z_u - z_T)$
\\
$\hat{S}$ & $1$ & 1 & 0 & 0 & $1$
\\
$\hat{T}$ & $\bar 3$ & 1 & 1/3 & 2/3 & $z_T$
\\
$\hat{\bar T}$ & $3$ & 1 & -1/3 & -2/3 & $-1-z_T$
\\
\hline
\end{tabular}
\end{center}
\label{charges}
\end{table*}%
\end{center}
\vspace{-0.5cm}
The most general superpotential that can be written with these charges is:
\begin{eqnarray}
\label{W.1}
	{\cal W}_1 &= & Y_t \, \hat{Q}_3 \, \hat{H}_u \, \hat{u}^c_3 \ + \  Y_b \, \hat{Q}_3 \, \hat{H}_d \, \hat{d}^c_3 \nonumber
\\
		& + & Y_\tau \, \hat{L}_3 \, \hat{H}_d \, \hat{e}^c_3 \ + \  Y_{\nu_3} \, \hat{L}_3 \, \hat{H}_u \, \hat{\nu}^c_3  \nonumber \\
	& + & \lambda \, \hat{S} \, \hat{H}_u \, \hat{H}_d + \lambda_1 \, \hat{S} \, \hat{T}  \, \hat{\bar T}  \nonumber \\
	& + & \lambda_2 \, \hat{Q}_3 \, \hat{L}_3 \, \hat{T} + \lambda_{3} \, \hat{u}^c_3 \, \hat{d}^c_3 \, \hat{T}  + \lambda_4  \hat{u}^c_3 \, \hat{e}^c_3 \, \hat{\bar T} \nonumber \\
	 & + & \lambda_{5} \, \hat{d}^c_3 \, \hat{\nu}^c_3 \,  \hat{\bar T} \ + \  \lambda_6 \hat{Q}_3 \hat{Q}_3 \hat{\bar{T}} .
\end{eqnarray}
The first and second rows allow for mass terms for the third generation only, the third for trilinear terms that, once $S$ acquires a VEV, generate mass terms for the MSSM Higgsinos and the colored exotics. 
The fourth and fifth rows sport third generation baryon and lepton number violating trilinear terms. These destabilize the proton but the lifetime can still be safe as it is suppressed by several CKM-like factors as will be discussed in a later section.  In addition, the typical MSSM non-renormalizable terms which violate baryon and lepton number are also allowed for the first and second generation.

This still leaves the first and second generation masses to be desired.  However, as can be appreciated from Table~\ref{charges}, there are still two degrees of freedom left: $z_T$ and $z_u$.  This allows a choice between tree-level down- or up-type quark masses.  We opt for tree-level up-type quark masses, which require $z_{H_u} = 0$ and yields the relationship $z_T = -1 - 2 z_u$ and new contributions to the superpotential of the form:
\begin{eqnarray}
	{\cal W}_2 & = & Y_u^{ab} \, \hat{Q}_a \, \hat{H}_u \, \hat{u}^c_b \ + \ Y_\nu^{ab} \, \hat{L}_a \, \hat{H}_u \, \hat{\nu}^c_b \nonumber \\
	 & + &  \lambda_d^{ab}  \frac{\hat{S}}{\Lambda} \hat{Q}_a \, \hat{H}_d \, \hat{d}^c_b \ + \  \lambda_e^{ab}  \frac{\hat{S}} {\Lambda} \hat{L}_a \, \hat{H}_d \, \hat{e}^c_b,
\end{eqnarray}
where $a,b = 1..2$ only.  In addition to the tree-level up-type masses, we can also generate down-type masses at the non-renormalizable level for the first and second generation.  
At this point, the only aspect of the fermionic sector missing is the mixings between the third generation and the others two.  Fortunately, we have yet another charge degree of freedom.

There are three possible scenarios that give CKM-like mixings: $z_T = -3, \ z_u = 1$; $z_T = 1, \ z_u = -1$; and $z_T=-1, \ z_u =0$.  Unfortunately, the latter two solutions introduce couplings between 
the colored exotic fields and the first two generations making proton decay unsafe.  This leaves the first solution as the unique realistic case with charges given by
\begin{eqnarray}
\hat{Q}_3 &\sim& -1, \hat{u}^c_3 \sim 1, \hat{d}^c_3 \sim 2, \hat{L}_3 \sim 4, \hat{e}^c_3 \sim -3, \nonumber \\ 
\hat{\nu}^c_3 & \sim & -4, \hat{H}_d \sim -1, \hat{H}_u \sim 0, \hat{S} \sim 1, \hat{T} \sim -3, \rm{and}\,  \hat{\bar T} \sim 2. \nonumber
\end{eqnarray}
Then, the additional superpotential terms allowed are
\begin{eqnarray}
	{\cal W}_3 &=& \lambda_7^a \frac{\hat{S}}{\Lambda} \hat{Q}_3 \, \hat{H}_u \, \hat{u}^c_a \ + \  \lambda_8^a \frac{\hat{S}^4}{\Lambda^4} \hat{L}_a \, \hat{H}_u \, \hat{\nu}^c_3  \nonumber \\
	& + & \lambda_9^a \frac{\hat{S}^2}{\Lambda^2} \hat{Q}_3 \, \hat{H}_d \, \hat{d}^c_a \ + \  \lambda_{10}^a \frac{\hat{S}^4}{\Lambda^4} \hat{L}_a \, \hat{H}_d \, \hat{e}^c_3,
\end{eqnarray}
where the first and third terms allow for a realistic CKM matrix while the second and fourth terms are relevant for the mixing matrix in the leptonic sector, the PMNS matrix, 
given the appropriate scale, $\Lambda$.  In order to generate the right value for the mass of strange quark ($ms(M_Z)\approx 56$ MeV, see for example Ref.~\cite{German}) 
we need a ratio, $\left<S \right>/\Lambda \approx 10^{-4}-10^{-3}$. This means one needs new degrees of freedom not very far from the TeV scale to understand the origin of these higher-dimensional operators. 
For example, one could integrate out some new fermions and generate the mass terms listed above. In this paper we will ignore the origin of these terms and consider an effective theory where we can understand 
the origin of the $\mu$-term and $M$-parity violating interactions.
%
%
\section{IV. Symmetry Breaking}
Symmetry breaking proceeds through the following VEVs: $\left<H_u^0\right> \equiv v_u/\sqrt{2}$ and $\left<H_d^0\right> \equiv v_d/\sqrt{2}$, responsible for EWSB; $\left< \tilde \nu^c \right> \equiv v_R/\sqrt 2$ (we will assume only one generation of right-handed sneutrinos acquires a VEV), breaking $B-L$~\cite{PRL}. The VEV $\left< S \right> \equiv v_S/\sqrt 2$ breaks $U(1)_S$, 
and $\left< \tilde \nu \right> \equiv v_{L}/\sqrt{2}$ is also generated.  Due to the non-universality of the $U(1)_S$ charges, the minimization conditions and Higgs spectrum depend 
on which generation of right-handed sneutrino acquires a VEV.  We will proceed in the most general way, designating the charges of the right-handed and left-handed sneutrino 
as $z_{\nu^c}$ and $z_{L}$, respectively. This is of course zero for the first two generations and $\mp 4$ for the third. We also elucidate the relevant soft parameters:
\begin{eqnarray}
	-\mathcal{L}_\text{Soft} &= & \left(a_\nu \tilde L \, H_u \, \tilde \nu^c + a_\lambda S \, H_u \, H_d + \text{h.c.}\right) \nonumber \\
	 & + & m_S^2 \left| S\right|^2 + m_{H_u}^2 \left| H_u\right|^2 + m_{H_d}^2 \left| H_d\right|^2 \nonumber \\
	 &+ & m_{\tilde{L}}^2 \left| \tilde{L} \right|^2 + m_{\tilde{\nu}^c}^2 \left| \tilde{\nu}^c \right|^2.
\end{eqnarray}
The VEVs of the potential in the phenomenologically appropriate limit of very small $v_L$ and $Y_\nu$ and in the one family approximation are
\begin{eqnarray}
	\nonumber
	\left<V_F\right> & = & -\frac{1}{2} Y_\nu \, \lambda \, v_L \, v_d \, v_R \, v_S \ + \ 
		\frac{1}{4} \lambda^2 \left(v_u^2 + v_d^2\right) v_S^2
\\
		& + & \frac{1}{4} \lambda^2 \, v_d^2 \, v_u^2,
	\\
	\nonumber
	\left<V_\text{S} \right> & = & \frac{1}{2} m_{H_u}^2 v_u^2 + \frac{1}{2} m_{H_d}^2 v_d^2
		+ \frac{1}{2} v_L^2 m_{\tilde{L}}^2 + \frac{1}{2} v_R^2 m_{\tilde{\nu}^c}^2
\\
		& + & \frac{1}{2} m_S^2 v_S^2
		+\frac{1}{\sqrt 2} a_\nu \, v_L \, v_u \, v_R - \frac{1}{\sqrt 2} a_\lambda \, v_d \, v_u \, v_S, \nonumber \\
\\
	\nonumber
	\left<V_D\right> & = & \frac{1}{32} \left(g_1^2 +g_2^2\right)\left(v_u^2 - v_d^2 - v_L^2\right)^2
\\
	\nonumber
		& + &
		\frac{1}{32} g_S^2 \left(v_S^2 -v_d^2 + z_{\nu^c} \, v_R^2  - z_{\nu^c} \, v_L^2\right)^2
\\
		 & + & 
		\frac{1}{32} g_{BL}^2 \left(v_R^2 - v_L^2\right)^2.
\end{eqnarray}
Focusing now on the scenario where the first or second generation sneutrinos acquire a VEV and 
assuming that $v_S,v_R \gg v_u, v_d$, so that the two sectors decouple, yields the following familiar MSSM-like results:
\begin{eqnarray}
	\frac{2 b}{\sin 2 \beta} &= & M_{H_u}^2 + M_{H_d}^2 + 2 \left| \mu \right|^2,
	\\
	\frac{1}{2} \ M_Z^2 &= & -\left| \mu \right|^2 - \left( \frac{M_{H_u}^2 \tan^2  \beta - M_{H_d}^2}{\tan^2 \beta -1}\right),
\end{eqnarray}
where the difference from the MSSM is in the definition of $M_{H_u}$ and $M_{H_d}$
\begin{eqnarray}
\label{newMSSM}
	M_{H_d}^2 & = & \, m_{H_d}^2 - \frac{1}{8} g_S^2 \left(v_S^2 - v^2 \cos^2 \beta \right)
		+\frac{1}{2} \lambda^2 v^2 \sin^2 \beta, \nonumber \\ &&
	\\
	M_{H_u}^2 & = & \, m_{H_u}^2 + \frac{1}{2} \lambda^2 v^2 \cos^2 \beta,
\end{eqnarray}
\begin{eqnarray}
	b & = & \frac{a_\lambda \ v_S}{\sqrt 2},
	\\
	\mu & = & \frac{1}{\sqrt 2} \lambda \ v_S.
\end{eqnarray}
Here, $v^2 \equiv v_u^2 + v_d^2$. The non-MSSM VEVs can be approximated as
\begin{eqnarray}
\label{VEV1}
	v_R^2 & = & - 8 \frac{m_{\tilde{\nu}^c}^2}{g_{BL}^2},
	\\
	v_S^2 & = & - \frac{\left(8 \, m_S^2 + 4 \lambda^2 v^2 - g_S^2 v^2 \cos^2 \beta\right)}{g_S^2},
	\label{VEV}
	\\
	v_L & = & \frac{v_R \left(\lambda Y_\nu \, v_d \, v_S - \sqrt{2} \, a_\nu v_u \right) }
			{2 \left(m_{\tilde L}^2 - \frac{1}{8} g_{BL}^2 v_R^2 +
			\frac{1}{8} \left(g_1^2 + g_2^2\right) v^2 \cos \left(2 \beta \right)\right)}.
\end{eqnarray}
Notice that using Eqs.~(17) and (19) one can understand that the $\mu$ term generated after symmetry breaking is determined by the soft mass $m_{S}$. Then, in this way one can say 
that SUSY breaking scale sets the size of this mass term in the MSSM superpotential.

The first two VEVs require the numerator to be positive meaning in general that $m_S^2, m_{\tilde{\nu}^c}^2 < 0$, \textit{i.e.} tachyonic right-handed sneutrino and singlet masses.  
A tachyonic $S$ can easily be generated through a radiative mechanism if its coupling to the exotic triplets is large enough, while --- for non-universal right-handed sneutrino masses --- a tachyonic right-handed sneutrino can be generated via the mechanism discussed in Ref.~\cite{Ambroso:2009jd}.  Alternatively, its possible that $\lambda_5$ is of order one, which will drive the right-handed sneutrino negative in the traditional way, however this would require much smaller values for the exotic triplet couplings to quarks to compensate for proton decay. Regardless of how the tachyonic masses are generated, 
the VEVs and therefore the symmetry breaking scales are defined by the SUSY breaking mass scale. This is very appealing since it tethers the corresponding $Z'$ masses to this 
scale as well, giving hope that the underlying mechanism for the $\mu$-term and $M$-parity violation can be tested at the LHC.
\section{V. Spectrum}
In this section we will outline the spectrum in the different sectors of this theory. 
\subsection{1. Charged Fermion Masses}
It is crucial to show that a consistent scenario for fermion masses is possible in this context.
A detailed analysis is beyond the scope of this article but a brief discussion is presented.
The charged fermion masses are generated after the symmetry breaking and are given by
\begin{equation}
{\cal M}_u=
\left(
\begin{array}{cc}
A_u & 0 \\
B_u & C_u
\end{array}
\right)
\end{equation}
where $A_u=Y_u v_u / \sqrt{2}$ is a 2 by 2 matrix, $B_u=\lambda_7 v_S v_u / 2 \Lambda$ is a 2 by 1 matrix, and $C_u=Y_t v_u / \sqrt{2}$. 
In the case of the down sector we find
\begin{equation}
{\cal M}_d=
\left(
\begin{array}{cc}
A_d & 0 \\
B_d & C_d
\end{array}
\right)
\end{equation}
where $A_d=\lambda_d v_s v_d / 2 \Lambda$ is a 2 by 2 matrix, $B_d=\lambda_9 v_S^2 v_d / 2 \sqrt{2} \Lambda^2$, and $C_d=Y_b v_d /  \sqrt{2}$. The mass matrix for charged leptons reads as
\begin{equation}
{\cal M}_e=
\left(
\begin{array}{cc}
A_e & 0 \\
B_e & C_e
\end{array}
\right)
\end{equation}
with $A_e=\lambda_e v_s v_d / 2 \Lambda$ is a 2 by 2 matrix, $B_e=\lambda_{10} v_S^4 v_d / 4 \sqrt{2} \Lambda^4$, and $C_e=Y_\tau v_d /  \sqrt{2}$.

There are two interesting results we should discuss: The first is that the new gauge symmetry $U(1)_S$ is basically a flavor symmetry since after symmetry breaking one obtains specific 
textures for all fermion mass matrices. Second, the fact that $B_{d,e}\ll A_{d,e} \ll C_{d,e}$ implies that the mass matrices for down-quark and charged leptons can be diagonalized approximately by a matrix containing a submatrix in the 2 by 2 sector. As we will explain carefully later, this ensures that the new physical couplings of the gauge boson associated to $U(1)_S$ will never induce large flavor violation is the down sector. This is important for avoiding the strong bounds from flavor changing neutral currents and proton decay. Notice that the same argument holds also for the up-quark sector, but with a less strong hierarchy. 
%
\subsection{2. Neutral Gauge Bosons}
We have two new neutral gauge bosons associated with the two new $U(1)$-groups. We proceed by assuming a sneutrino VEV in the first or second generation only and no kinetic mixing terms.  The mass matrix for the four neutral gauge bosons in the basis $(B^{Y}_\mu,W_{\mu}^3,B^{S}_\mu,B^{BL}_\mu)$ is then
\begin{eqnarray}
&&
\left(
\begin{array}{cccc}
	\frac{1}{4} g_1^2 v^2
	& -\frac{1}{4} g_1 g_2 v^2
	&  \frac{1}{4}g_1 g_S v_d^2
	& 0
	\\
	-\frac{1}{4} g_1 g_2 v^2
	& \frac{1}{4} g_2^2 v^2
	& -\frac{1}{4}g_2 g_S v_d^2
	& 0
	\\
	\frac{1}{4}g_1 g_S v_d^2
	& -\frac{1}{4}g_2 g_S v_d^2
	& \frac{1}{4}g_S^2 \left(v_d^2+v_S^2\right)
	&  0
	\\
	0
	& 0
	& 0
	& \frac{1}{4}g_{BL}^2 v_R^2
\end{array}
\right), \nonumber \\ &&
\end{eqnarray}
where $v^2\equiv v_u^2 + v_d^2\approx (246~\mbox{GeV})^2$. Thus we can immediately see that the $B^{BL}_\mu$ gauge boson does not mix with the other neutral gauge bosons and decouples.  We define the mass eigenstate as $Z_{BL}$ with mass $\frac{1}{4}g_{BL}^2 v_R^2$. Rotating by the weak angle $\theta_W$ projects out the photon zero-mode which decouples and leaves the two-by-two mass matrix in the basis $(Z^{0}_\mu,B^{S}_\mu)$
\begin{eqnarray}
\mathcal{M}_{ZZ^{'}}^2 =
\left(
\begin{array}{ccc}
	 M^2_{Z^0}
	 & \Delta
	 \\
	 \Delta
	 & M^2_{Z_S}
\end{array}
\right),
\end{eqnarray}
where
\ba
	M^2_{Z^{0}}	&=& \frac{1}{4} v^2 \left(g_1^2 + g_2^2\right),		\\
	M^2_{Z_S}	&=& \frac{1}{4}g_S^2   \left(v_d^2 + v_S^2\right),		\\
	\Delta &=& -\frac{1}{4}v_d^2  g_S \sqrt{g_1^2+g_2^2} \nonumber \\
	&=& -\frac{g_S }{\sqrt{g_1^2+g_2^2}}M_{Z^0}^2 \cos^2 \beta~.
\ea
This matrix describes the $Z^{0}_\mu$-$B^{S}_\mu$ mixing. The non-diagonal element is proportional to $v_d$ and thus the mixing will be suppressed for large values of $\tan{\beta}$. We label the physical states $Z$ and $Z'$ whose masses are
\beq
M^2_{Z,Z'}=\frac{1}{2} \left[  M_{Z^0}^2+ M_{Z_S}^2\mp \sqrt{(M_{Z^0}^2- M_{Z_S}^2 )^2+ 4 \Delta^2} \right]~,
\eeq
which in the limit $M_{Z^0}^2 \ll M_{Z_S}^2 $ simplifies to
\begin{eqnarray}
	M^2_Z & \approx & M^2_{Z^{0}} + \frac{g_S^2}{g_1^2 + g_2^2} \frac{M_{Z^0}^4 \cos^4 \beta}{M_{Z^0}^2 - M_{Z_S}^2}, 
	\\
	M^2_{Z'}	& \approx & M^2_{Z_{S}} - \frac{g_S^2}{g_1^2 + g_2^2} \frac{M_{Z^0}^4 \cos^4 \beta}{M_{Z^0}^2 - M_{Z_S}^2}.
\end{eqnarray}
The mixing angle, defined such that
\begin{eqnarray}
	Z^0_\mu & =  Z_\mu \cos \theta_{ZZ'} - Z_\mu' \sin \theta_{ZZ'},
	\\
	{Z_S}_\mu & =  Z_\mu \sin \theta_{ZZ'} + Z_\mu' \cos \theta_{ZZ'},
\end{eqnarray}
\begin{eqnarray}
\label{theta.epsilon}
	\theta_{ZZ'} &=& \frac{1}{2}  \arctan \left( \frac{2\Delta}{M_{Z^0}^2- M_{Z_S}^2} \right) \nonumber \\
	&\approx & \frac{g_S}{\sqrt{g_2^2+g_1^2}}\cos^2{\beta}\,\epsilon + {\cal O} (\epsilon^2).
\end{eqnarray}
Here $\epsilon \equiv \frac{M_{Z^0}^2}{M_{Z_S}^2}$. Notice that the ${\cal O} (\epsilon)$-terms in $M^2_{Z, Z'}$ and $\theta_{ZZ'}$ have an additional 
suppression for large values of $\tan{\beta}$.  For a recent discussion of the constraints on $\theta_{ZZ'}$ see Ref.~\cite{Erler-Langacker}.

In order to illustrate the possible numerical values for the mixing angle $\theta_{ZZ'}$ in Fig. 1 we show the values when $v_S=2$ TeV 
and for different values of $g_S$ and $\tan \beta$. Notice that in the whole parameter space the mixing angle is very small, i.e. $\theta_{ZZ'} < 10^{-3}$.
\begin{figure*}[t!]
	\plot{0.5\linewidth}{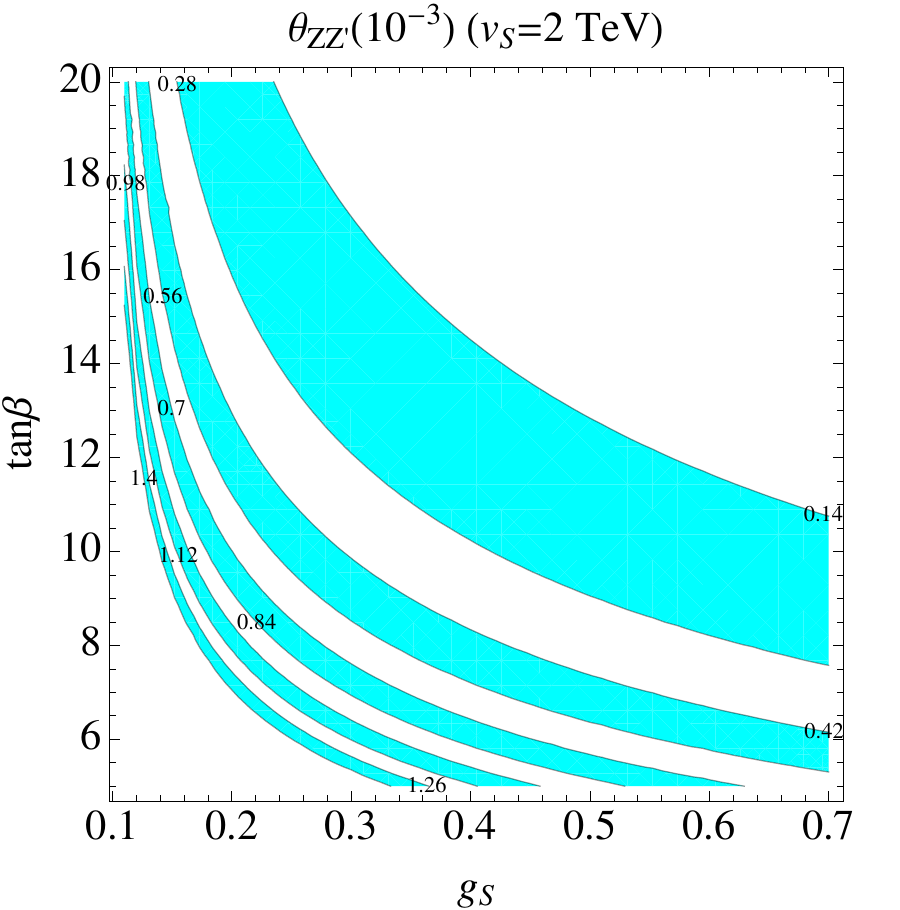} 
	\caption{Values for the mixing angle $\theta_{ZZ^{'}}$ for different values of $\tan\beta$ and the gauge coupling $g_S$ when $v_S=2$ TeV.}
\end{figure*}
Before going to the next subsection, let us make some comments about the case where the third generation right-handed sneutrino acquires a nonzero VEV. In that case we have the following Z-mass matrix
\begin{eqnarray}
&\left(
\begin{array}{ccc}
	\frac{1}{4} v^2 \left(g_1^2+g_1^2\right)
	& -\frac{1}{4}v_d^2   g_S \sqrt{g_2^2+g_1^2}
	& 0
	\\
	-\frac{1}{4}v_d^2 g_S \sqrt{g_2^2+g_1^2}
	& \frac{1}{4}g_S^2 \left(v_d^2+v_s^2+16 v_R^2\right)
	& -g_{BL} g_S v_R^2
	\\
	0
	& -g_{BL} g_S v_R^2	
	&	\frac{1}{4}g_{\text{BL}}^2 v_R^2
\end{array}
\right). \nonumber
&
\end{eqnarray}
This is a more complicated case since the two new $Z'$ bosons do mix, and they also mix with the SM Z-boson. If one of the new Z-bosons is much heavier than the other, then it decouples and we are in the usual Z-Z' scenario, whereas if both have similar masses then one has a Z-Z'-Z'' situation where the expressions are more involved (see Ref.~\cite{Heeck:2011md} for an analysis of the kinetic and mass mixing of three neutral gauge bosons). In any case, the mixing of the heavy states with the SM Z-boson and the contribution to the SM Z-mass are still dominated by the quantity $\frac{v^2}{v^2_{S,R}}\cos{\beta}$ in such a way that, as in the previous case, they are suppressed for large values of $\tan{\beta}$ and $v_{S,R}$.

The phenomenology of a $B-L$ gauge boson has been studied extensively in the literature and relevant bounds can be found in Ref.~\cite{Carena:2004xs}.  The reach at the LHC for a $B-L \ Z'$ is studied in Ref.~\cite{Basso:2010pe} and the effects of SUSY decays are shown in Ref.~\cite{Chang:2011be}.  While a $Z'$ that couples only to the third family does not have as much coverage it has been studied in Ref.~\cite{Andrianov:1998hx}.
{\subsection{3. $Z^{'}$ Couplings to Fermions}}
Here we study the case where only the $U(1)_S$ $Z'$ and the SM $Z$ boson mix. The neutral current interactions of the fermions are described by the Lagrangian
\ba
- {\cal L}_{Z^{'}}
&=& g_1 J_Y^\mu B_{\mu} + g_2 J_3^\mu W_{\mu}^3 + g_S J_S^\mu B^S_{\mu} + g_{BL} J_{BL}^\mu B^{BL}_{\mu}	\nonumber \\
&=& e J_{em}^\mu A_{\mu} + g_0 J_0^\mu Z^{(0)}_{\mu} + g_S J_S^\mu B^S_{\mu} + g_{BL} J_{BL}^\mu B^{BL}_{\mu}	 \nonumber \\
&=& e J_{em}^\mu A_{\mu} + g_Z J_Z^\mu Z_{\mu} + g_{Z'} J_{Z'}^\mu Z'_{\mu} + g_{BL} J_{BL}^\mu Z^{BL}_{\mu}, \nonumber \\
\ea
where $J_3^\mu, J_Y^\mu,J_{em}^\mu,J_0^\mu$ are the well known SM currents. The electromagnetic and $B-L$ currents are not modified whereas $J_Z^\mu$ and $J_{Z'}^\mu$ are
\ba
\label{eq:NCZ}
J_Z^\mu	&=&	\bar{u} \gamma^\mu \left( C_{uL} P_L + C_{uR} P_R \right) u  \nonumber \\ 
&+ & \bar{d} \gamma^\mu \left( C_{dL} P_L + C_{dR} P_R \right) d, \\
\label{eq:NCZp}
J_{Z'}^\mu	&=&	\bar{u} \gamma^\mu \left( C'_{uL} P_L + C'_{uR} P_R \right) u \nonumber \\
&+ & \bar{d} \gamma^\mu \left( C'_{dL} P_L + C'_{dR} P_R \right) d~,
\ea
where $u^T\equiv (u,c,t)$, $d^T\equiv (d,s,b)$ and the $C$ matrices are three-by-three charge matrices in flavor space. 
The currents $J_0^\mu$ and $J_S^\mu$ have the same structure, and the relation between the $C$-matrices in the different bases is the following
\ba
g_Z C_{x}		&=&	g_0 C_{x}^{0} \cos{\theta_{ZZ'}} + g_S C_{x}^{S} \sin{\theta_{ZZ'}},		\\
g_{Z'} C'_{x}		&=&	- g_0 C_{x}^{0} \sin{\theta_{ZZ'}} + g_S C_{x}^{S} \cos{\theta_{ZZ'}}~,
\ea
where $x=uL,uR,dL,dR$. The $C_{x}^{0}$ matrices are those of the SM and are proportional to the identity (flavor universal interaction), whereas the $C_{x}^{S}$ matrices are non-universal because only the third generation feels the $U(1)_S$ interaction.
So far we have only taken into account the effects of the EWSB in the gauge sector, with the associated mixing among $Z$ bosons, but mixing in the fermion sector must also be taken into account. Starting with the Yukawa matrices that have been introduced in the previous sections and performing the usual rotation to mass-eigenstates:
\ba
&&	u_{L,R}		\to	U_{L,R}~u_{L,R},	\\
&&	d_{L,R}		\to	D_{L,R}~d_{L,R},
\ea
we end up with a Lagrangian with the usual CKM matrix in the charged current sector $V_{CKM}=U_L^\dagger D_L$ (we neglect possible extra phases). 
In the neutral current sector we have the same structures \eqref{eq:NCZ} and \eqref{eq:NCZp}, but making the substitutions
\ba
C_{uL}		\to		\tilde{C}_{uL} \equiv U_{L}^\dagger C_{uL} U_{L} ~,
\ea
and the same transformation holds for $C_{uR},C_{dL}$ and $C_{dR}$, and for the $Z'$-current. 
As it is well-known, the SM $C_{x}^{0}$ matrices remain unchanged by this rotation because they are proportional to the identity, but things are different for the $C_{x}^{S}$ matrices, where we will have
\ba
\tilde{C}_{dL}^{S}
&=&	D_L^\dagger C_{dL}^{S} D_L \approx C_{dL}^{S}~,	\\
\tilde{C}_{dR}^{S}
&=&	D_R^\dagger C_{dR}^{S} D_R \approx C_{dR}^{S}~,\\
\tilde{C}_{uL}^{S}
&=&	U_L^\dagger C_{uL}^{S} U_L \approx V_{CKM} C_{uL}^{S} V^\dagger_{CKM}~,	\\
\tilde{C}_{uR}^{S}
&=&	U_R^\dagger C_{uR}^{S} U_R~.
\ea
where we have neglected the non-diagonal elements in the Yukawa couplings in the down sector involving the third family and that the $C^S_{x}$ matrices are zero except for the (3,3) element. Thus one can see that, apart from the very small mixing, the only new effect in the coupling of the Z-boson to the down quarks is in the diagonal $Zb\bar{b}$ coupling, which will be slightly modified. We have only discussed things in the quark sector, but the leptonic sector is identical.

In the up-quark sector things are different and FCNC are in principle possible  
\ba
\left[\tilde{C}^{S}_{uL}\right]_{ij} &=&	(V_{CKM})_{i3} \left[C_{uL}^{S}\right]_{33} (V_{CKM})^*_{j3} \\
&=& - (V_{CKM})_{i3} (V_{CKM})^*_{j3}~,	\\
\left[\tilde{C}^{S}_{uR}\right]_{ij} &=&	\left[U_R\right]^*_{3i} \left[C_{uR}^{S}\right]_{33} \left[U_{R}\right]_{3j} = - \left[U_R\right]^*_{3i} \left[U_{R}\right]_{3j}. \nonumber \\
\ea
However, the argument given for the down-quark sector can be applied also here in the limit where we neglect the higher dimensional operators. Therefore the FCNC in the up-quark sector are suppressed by the smallness of the elements $\left[M_u\right]_{i3}$ and $\left[M_u\right]_{3i}$ ($i=1,2$) in our model, although in the left-handed sector this is related to the CKM matrix.
Thus we see that FCNC in the down-quark and charged-lepton sector, where the strongest constraints appear ($K^0-\bar{K}^0$ mixing, $\mu-e$ conversion, ...) \cite{Langacker-Plumacher} 
are suppressed in our model. In the up-quark sector we have found that the FCNC are suppressed by the tiny $\left[M_u\right]_{i3,3i}$ elements and also either by the small mixing $\theta_{ZZ'}$, or by the mass of the $Z'$ boson. One can actually check that the Yukawa suppression is so strong that even for a $Z'$ boson lighter than the SM $Z$ boson one satisfies the constraints coming from $D^0-\bar{D}^0$ mixing~\cite{Golowich}.
A detailed analysis of all the constraints coming from flavor violation will be published in a future publication.
\subsection{4. Higgs Sector}
The Higgs sector is composed of the MSSM Higgs doublets, $H_u$ and $H_d$, and the singlet $S$. After symmetry breaking lepton number is broken and the Higgses will mix with the sneutrinos in the theory.
Realistic neutrinos masses constrain this mixing to be quite small hence decoupling the left-handed sleptons from the Higgs bosons (although these effects can be important in the decays of the LSP). 
Keeping this in mind, the physical Higgs sector contains one CP-odd scalar $A$, similar to the MSSM but now with some small admixture of $S$.  It also contains four CP-even scalars: $h$, the SM-like Higgs; and $H_1, H_2$ and $H_3$.  
The latter three are some combination of the MSSM Higgs bosons, $S$ and the right-handed sneutrino.  These are labeled from lightest to heaviest.  Of course, there is also the charged Higgs of the MSSM, 
$H^\pm$, whose composition is purely MSSM Higgs bosons.

The mass of the CP-odd Higgs is given by
\begin{equation}
\label{mass.A}
	m_A^2  = \frac{2 b}{\sin 2 \beta} + \frac{b \,  v^2 \sin 2 \beta}{2 v_S^2}.
\end{equation}
There are two limits in which the $Z-Z'$ mixing is phenomenologically viable: $M_{Z^0}^2/M_{Z_S}^2 \ll 1$ (Eq.~(\ref{theta.epsilon})) which implies $v^2/v_S^2 \ll 1$ and when $\tan \beta$ is quite large.  Both cases imply the second term in the $m_A^2$ expression is negligible therefore yielding $m_A \sim 2\,b/\sin 2 \beta$ as in the MSSM.  This value is always positive.  The goldstone boson associated with the $U(1)_S$ and $U(1)_{B-L}$ will predominately be composed of a linear combination of the CP-odd part of $S$ and $\tilde{\nu}^c$ depending on the kinetic mixing between those two sectors and which generation of right-handed sneutrino acquires a VEV.
\\
The most general mass matrix for the CP-even scalars, $\mathcal M_S^2$, in the basis $\sqrt 2 \text{Re} \left(H_d, \ H_u, \ S, \  \tilde{\nu}^c\right)$, has the following elements:
\begin{eqnarray}
	\mathcal M_{S_{11}}^2 &=  & \frac{1}{4} \left(g_1^2 + g_2^2 + g_S^2\right) v^2 \cos^2 \beta + b \tan \beta,
	\\
	\mathcal M_{S_{12}}^2 & = & -b + \frac{1}{8}\left(4\lambda^2 - g_1^2 - g_2^2\right) v^2 \sin 2 \beta,	
\\
	\mathcal M_{S_{13}}^2 & = & \frac{1}{4}\left(4 \lambda^2 - g_S^2 \right) v \, v_S \cos \beta - b \frac{v}{v_S} \sin \beta,
\end{eqnarray}	
\begin{eqnarray}	
	\mathcal M_{S_{14}}^2 &= & - \frac{1}{4} g_S \left(\xi \, g_{BL} + z_{\nu^c} \, g_S\right) v_R \, v \cos \beta,
\\
	\mathcal M_{S_{22}}^2 &=  & \frac{1}{4} \left(g_1^2 + g_2^2\right) v^2 \sin^2 \beta  + b \cot \beta,
	\\
	\mathcal M_{S_{23}}^2 &=  & -b \frac{v}{v_S} \cos \beta + \lambda^2 v_S \, v \sin \beta,
	\\
	\mathcal M_{S_{24}}^2 & =  & 0,
	\\
	\mathcal M_{S_{33}}^2 & =  & \frac{1}{4} \, g_S^2 \, v_S^2 + \frac{1}{2} \, b  \frac{v^2}{v_S^2} \sin 2 \beta,
	\\
	\mathcal M_{S_{34}}^2 & =  & \frac{1}{4} g_S \left(\xi \, g_{BL} + z_{\nu^c} \, g_S\right) v_R \, v_S,
	\\
	\mathcal M_{S_{44}}^2 & =  & \frac{1}{4} \left(z_{\nu^c}^2 g_S^2 + 2 \, \xi \, z_{\nu^c} g_S \, g_{BL} + g_{BL}^2 \right) v_R^2,
\end{eqnarray}
where $\xi$ is the kinetic mixing between $U(1)_{B-L}$ and $U(1)_S$ and $z_{\nu^c}$ is the $U(1)_S$ charge of $\nu^c$: zero for the first two generations and negative four for the third.  In the case where these two parameters are zero, the right-handed sneutrino has a mass equal to the $Z_{BL}$ mass: $g_{BL} v_R/2$.
For completeness, we also present the important one-loop corrections, \cite{Degrassi:2009yq}, to the upper-left three-by-three matrix from top/stop loops presented in Ref.~\cite{NMSSMreview} and repeated here only for the sake of consistent notation.  Some of these are implemented by the redefinition in the tree-level mass matrix:
\begin{equation}
	b \to b + \frac{3}{16 \sqrt 2 \, \pi^2} \lambda Y_t^2 A_t F_t v_S
\end{equation}
while the rest are given by
\ba
	\Delta \mathcal M_{S_{11}}^2 &=  & - \frac{3 Y_t^2}{32 \pi^2} \mu^2 \ G_t,
\\
	\Delta \mathcal M_{S_{22}}^2 & =  & \frac{3 Y_t^2}{32 \pi^2} \times \\ 
	&& \left(  - A_t^2 \ G_t \ + \  4 A_t \ E_t \ + \  4 m_t^2  \  \rm{ln} \left(  \frac{M_{\tilde{t}_1}^2 M_{\tilde{t}_2}^2}{m_{t}^4}\right)  \right),   \nonumber 
\ea	
\ba
	\Delta \mathcal M_{S_{33}}^2 & =  & - \frac{3 Y_t^2}{64 \pi^2}  \lambda^2 v^2 \cos^2\beta \ G_t,
\\
	\Delta \mathcal M_{S_{12}}^2 & =  &  \frac{3 Y_t^2}{32 \pi^2} \mu  \left(  A_t \ G_t \ - \ 2 E_t  \right),
\\
	\Delta \mathcal M_{S_{13}}^2 & =  &  \frac{3 Y_t^2}{32\sqrt{2} \pi^2} \lambda \ \mu \ v  \  \cos \beta \left(  4 F_t - G_t \right),
\\
	\Delta \mathcal M_{S_{23}}^2 & =  & \frac{3 Y_t^2}{32\sqrt{2} \pi^2} \lambda \ v \  \cos \beta \left( A_t G_t \ - \ 2 E_t \right),	
\end{eqnarray}
where $m_t$ is the top mass, $M_{\tilde t_1}$ and $M_{\tilde t_2}$ are the lighter and heavier stop masses respectively and $A_t$ is the trilinear-$a$ term for the up-type Higgs and stops: 
$V_\text{Soft} \supset Y_t \, A_t \, \tilde Q \, H_u \, \tilde t^c$.  Finally, the loop functions are given by

\begin{eqnarray}
	F_t & = & \frac{1}{M_{\tilde t_2}^2 - M_{\tilde t_1}^2}
		\left(
			M^2_{\tilde t_2} \ln{\frac{M^2_{\tilde t_2}}{M_\text{SUSY}^2}}
			- M^2_{\tilde t_1} \ln{\frac{M^2_{\tilde t_1}}{M_\text{SUSY}^2}}
		\right)
		-1, \nonumber \\
\end{eqnarray}
\begin{eqnarray}
	G_t &= & \sin^2 2 \theta_{\tilde t}
	\left(
		\frac{M_{\tilde t_2}^2 + M_{\tilde t_1}^2}{M_{\tilde t_2}^2 - M_{\tilde t_1}^2}
		\ln \frac{M^2_{\tilde t_2}}{M^2_{\tilde t_1}}
		-2
	\right),
\\
	E_t & = & -m_t \sin 2 \theta_{\tilde t} \ln \frac{M^2_{\tilde t_2}}{M^2_{\tilde t_1}},
\end{eqnarray}
%
where $\theta_{\tilde t}$ is the mixing angle in the stop sector and $M_\text{SUSY}$ is typically taken to be $\sqrt{M_{\tilde t_1} M_{\tilde t_2}}$.
The physical stop masses are derived by diagonalizing the stop mass matrix:
\begin{equation}
	\mathcal M_{\tilde t}^2 = 
	\begin{pmatrix}
		m_{\tilde Q}^2 + m_t^2 + D_L
		&
		m_t X_t
		\\
		m_t X_t
		&
		m_{\tilde t^c}^2 + m_t^2  + D_R
	\end{pmatrix},
\end{equation}
where 
\ba
D_L &=& \left(\frac{1}{2} - \frac{2}{3} \sin^2 \theta_W\right) M_{Z}^2 \cos 2 \beta, \\
D_R &=& \frac{2}{3} \sin^2 \theta_W M_Z^2 \cos 2 \beta, \\
X_t &=& A_t - \mu \cot \beta. 
\ea  
The radiative correction to the Higgs mass is maximized for maximal mixing, defined as $X_t = \sqrt 6 M_{S}$, where $M_{S}^2 \equiv \frac{1}{2} \left(M_{\tilde t_1}^2 + M_{\tilde t_2}^2\right)$ and we use notation similar to Ref.~\cite{Carena:2002es}.

\begin{figure*}[t]
\centering
\begin{tabular}{cc}
	\plot{.50\linewidth}{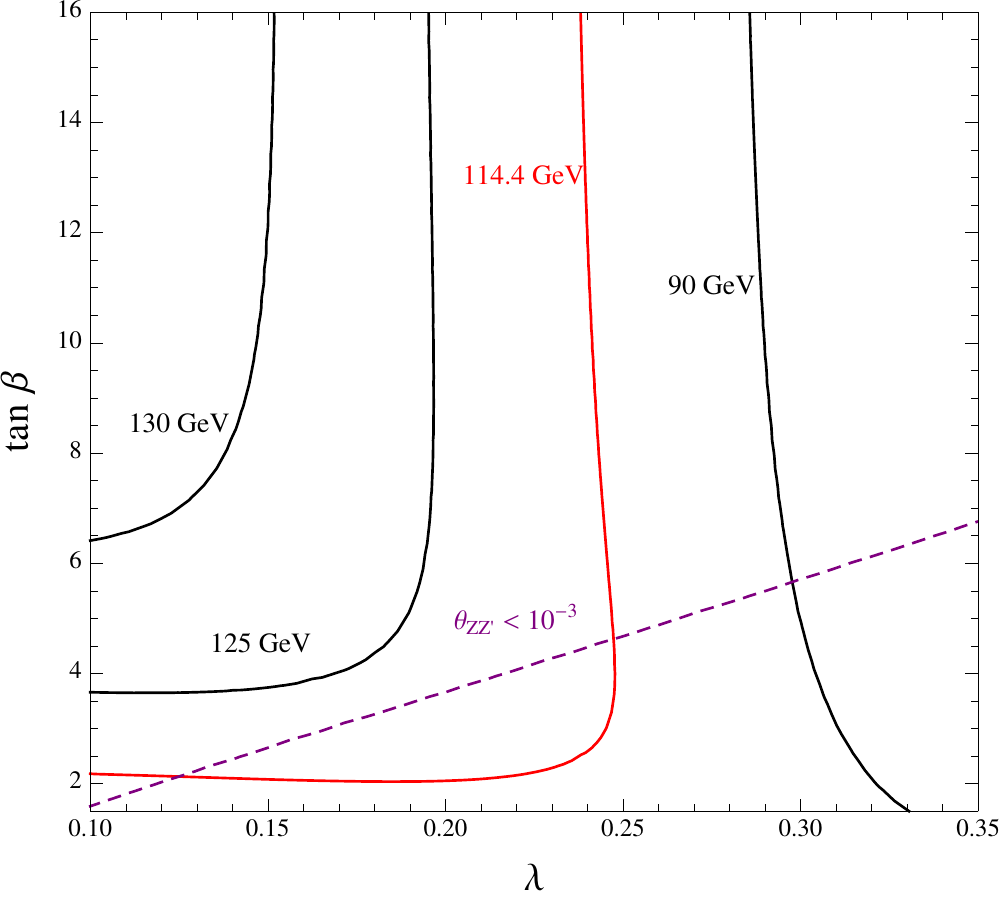} &
	\plot{.51\linewidth}{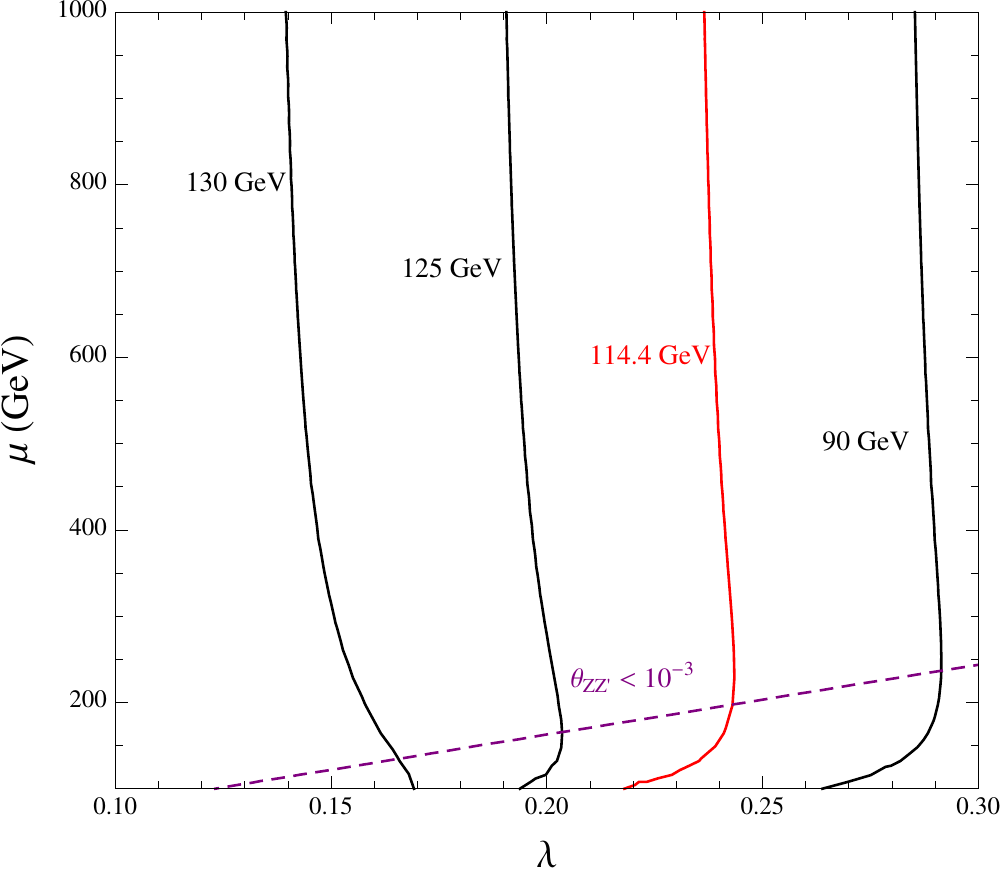}
	\\
	\put(40,5){\small (a)} &
	\put(40,5){\small (b)}
\end{tabular}
	\plot{.50\linewidth}{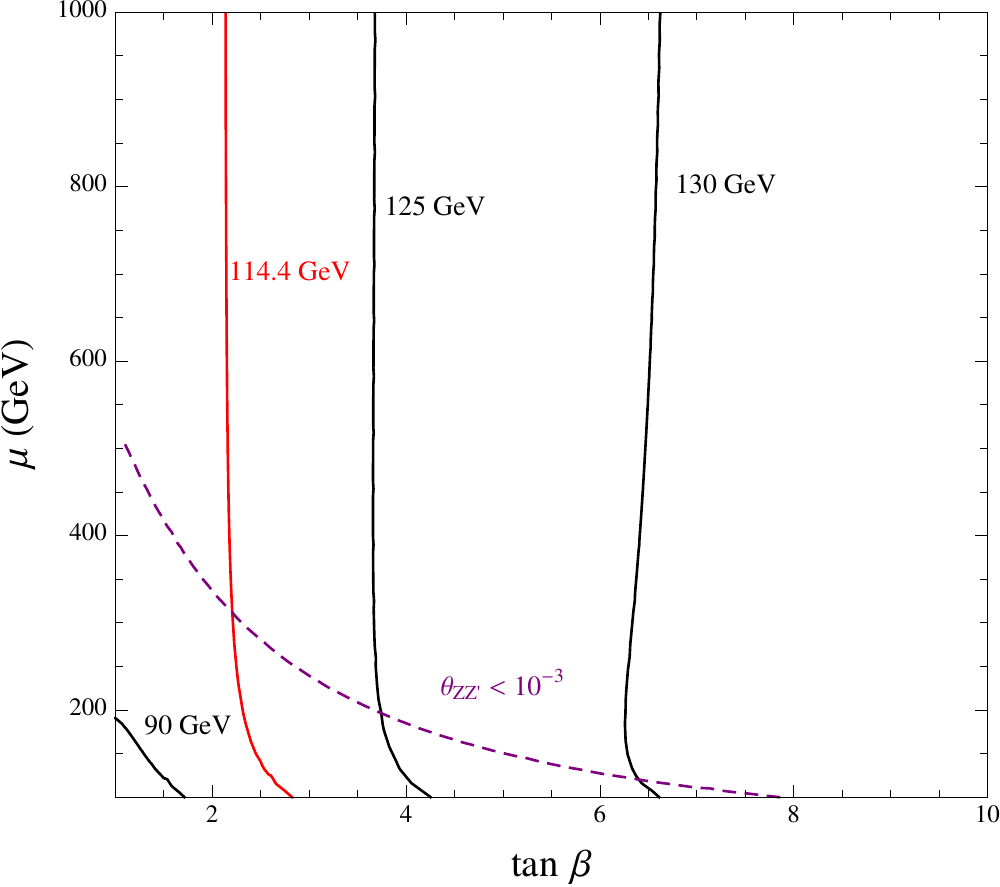}
	\put(-40,-4){\small (c)}
	\caption{Curves of constant $m_h$ in the (a) $\tan \beta - \lambda$ plane for $\mu = 400$ GeV, (b) $\mu - \lambda$ plane for $\tan \beta = 10$ and (c) $\mu - \tan \beta$ plane for $\lambda = 0.1$.  We use $a_\lambda = 100$ GeV, $g_S = 0.4$ and a top mass of $173$ GeV.  Here $B-L$ is broken by the second generation sneutrino only and there is no mixing between the $B-L$ and $U(1)_S$ sectors.  We further assume the maximal mixing scenario for the stop masses ($X_t = \sqrt 6 M_S$) with the soft mass parameters $m_{\tilde Q} = m_{\tilde t^c} = 1$ TeV.  The red contour is the LEP2 bound on $m_h$ of 114.4 GeV and the dashed purple lines indicate constant $\theta_{ZZ'} = 10^{-3}$: a conservative upper bound on the $Z-Z'$ mixing.}
\label{higgs.mass}
\end{figure*}

The SM-like Higgs mass will depend on the various parameters and the one-loop effects.  In Fig.~\ref{higgs.mass} we plot curves of constant $m_h$ in the (a)  $\tan \beta - \lambda$ plane for $\mu = 400$ GeV, (b) $\mu - \lambda$ plane for $\tan \beta = 10$ and (c) $\mu - \tan \beta$ plane for $\lambda =0.1$; the red curves correspond to the LEP2 bound of 114 GeV.  We furthermore use $a_\lambda = 100$ GeV, $g_S = 0.4$ and a top mass of $173$ GeV.  Dashed purple curves of constant $\theta_{ZZ'} = 1\times10^{-3}$ are also included as a conservative upper bound.  This calculation is done in the maximal mixing scenario ($X_t = \sqrt 6 M_{S}$), for $m_{\tilde Q} = m_{\tilde t^c} = 1000$ GeV.  This corresponds to $m_{\tilde t_{1,2}} \sim 800, \ 1180$ GeV.  We further assume no mixing between the $B-L$ and $U(1)_S$ sectors, \textit{i.e.} no kinetic mixing and no VEV for the third generation sneutrino.
Varying $a_\lambda$ also has an effect the contours, namely elongating the corners of the curves in (a) in (b) towards the right and in (c) towards the left but does not influence the maximum Higgs mass value.

Fig.~\ref{higgs.mass} indicates that the Higgs mass is maximized for small $\lambda$ and large $\tan \beta$.  Small $\lambda$ is one of the necessary limits to recover the MSSM, while increased Higgs mass with increased $\tan \beta$ is a behavior shared with the MSSM.  In fact, in both cases, the maximum is at around $m_h \sim 130$ GeV for this value of the stop masses and stop mixing.  The reason for the strong resemblance to the MSSM is that the NMSSM-like parameter space that allows for a Higgs mass surpassing the MSSM value---large $\lambda$, relatively small $\mu$ and small $\tan \beta$---is ruled out here due to $\theta_{ZZ'}$, see Fig.~\ref{higgs.mass}.  However it might be possible to relax this bound on $\theta_{ZZ'}$ since $Z_S$ couples only to the third generation.  While a more detailed study of this is required, this part of parameter space could open up new NMSSM-like possibilities such as the lightest Higgs being mostly singlet thereby pushing up the mass of the SM-like Higgs.  Since the mostly singlet Higgs and $Z_S$ have correlated masses, this would further mean a light $Z_S$ which could alleviate a tension that usually exists in models with gauge origins for the $\mu$ term: a tension between requiring a large $v_S$ for a large $Z'$ mass and a small $v_s$ for a small $\mu$ term required for reduced fine-tuning since.  We save further speculations for a future work.

Finally, the mass of the charged Higgs is
\begin{equation}
	m_{H^\pm}^2 = \frac{2 b}{\sin 2 \beta} + M_W^2 - \frac{1}{2} \lambda^2 v^2,
\end{equation}
where $M_W$ is the mass of the $W$ boson of the SM and where in general the above expression could be negative but will typically be dominated by the positive contribution from the $b$-term.
\vspace{-0.5cm}
\subsection{5. $M$-Parity Violation, Neutralinos and Neutrinos}
Above the SUSY scale, the $B-L$ symmetry guarantees $M$-parity conservation.  Once the right-handed sneutrino acquires a VEV bilinear $M$-parity violating terms (which break lepton number) are generated.  Schematically, these include $Y_\nu v_R \left(L H_u\right)$, the effective $\epsilon$ term and the only significant contribution from the superpotential, and gaugino-lepton mixing, \textit{e.g.}
\begin{align}
	g_{BL} v_R \left(\nu^c \tilde B_{BL}\right), \ g_2 v_L \left(\nu \tilde W^0\right), g_2 v_L \left(e \tilde W^+\right).
\end{align}
In addition to mediating the decay of the LSP, these terms mix SUSY and SM particles contributing to the neutralino mass matrix.
In the basis $\left(\nu, \nu^c, \tilde B_{BL}, \tilde S, \tilde B_S, \tilde B, \tilde W, \tilde H_d, \tilde H_u\right)$ the neutralino mass matrix is given by
\begin{equation}
\mathcal{M}_{\chi^0} =
\begin{pmatrix}
	M_\text{B-L}
	&
	\Gamma
	\\
	\Gamma^T
	&
	M_{\chi^0}
\end{pmatrix},
\end{equation}
where
\ba
M_\text{B-L} & = &
\begin{pmatrix}
	0
	&
	\frac{Y_\nu v_u}{\sqrt 2}
	&
	-\frac{g_{BL} v_L}{2}
\\
	\frac{Y_\nu v_u}{\sqrt 2}
	&
	0
	&
	\frac{g_{BL} v_R}{2}
\\
	-\frac{g_{BL} v_L}{2}
	&
	\frac{g_{BL} v_R}{2}
	&
	M_{BL}
\end{pmatrix},
\\ && \nonumber \\
\Gamma & = &
\begin{pmatrix}
0
	&
	\frac{g_S z_L v_L}{2}
	&
	-\frac{g_1 v_L}{2}
	&
	\frac{g_2 v_L}{2}
	&
	0
	&
	\frac{Y_\nu v_R}{\sqrt 2}
\\
	0
	&
	\frac{g_S z_\nu v_R}{2}
	&
	0
	&
	0
	&
	0
	&
	\frac{Y_\nu v_L}{\sqrt 2}
\\
	0
	&
	0
	&
	0
	&
	0
	&
	0
	&
	0
\end{pmatrix},
\ea
and
\ba
M_{\chi^0} &=&
\begin{pmatrix}
	0
	&
	\frac{g_S v_s}{2}
	&
	0
	&
	0
	&
	-\frac{\lambda v_u}{\sqrt 2}
	&
	-\frac{\lambda v_d}{\sqrt 2}
\\
	\frac{g_S v_s}{2}
	&
	M_S
	&
	0
	&
	0
	&
	-\frac{g_S v_d}{2}
	&
	0
\\
	0
	&
	0
	&
	M_1
	&
	0
	&
	-\frac{g_1 v_d}{2}
	&
	\frac{g_1 v_u}{2}
\\
	0
	&
	0
	&
	0
	&
	M_2
	&
	\frac{g_2 v_d}{2}
	&
	-\frac{g_2 v_u}{2}
\\
	-\frac{\lambda v_u}{\sqrt 2}
	&
	-\frac{g_S v_d}{2}
	&
	-\frac{g_1 v_d}{2}
	&
	\frac{g_2 v_d}{2}
	&
	0
	&
	-\frac{\lambda v_s}{\sqrt 2}
\\
	-\frac{\lambda v_d}{\sqrt 2}
	&
	0
	&
	\frac{g_1 v_u}{2}
	&
	-\frac{g_2 v_u}{2}
	&
	-\frac{\lambda v_s}{\sqrt 2}
	&
	0
\end{pmatrix}, \nonumber \\
\ea
where the lower four-by-four block of $M_{\chi^0}$ is the MSSM mass matrix and $M_{\text{B-L}}$ is the $B-L$ part and decouples from the rest if the third generation sneutrino is not VEVed and there is no significant $(B-L)$--$S$ mixing.  If only one sneutrino acquires a VEV, there will be one heavy right-handed neutrino and two with active neutrino masses~\cite{Ghosh:2010hy}.

In the case of a third generation right-handed sneutrino VEV, an additional $M$-parity violating term is generated, which mixes the right-handed bottom quark (squark) with the triplino (triplet) via the $\lambda_5$ couplings in Eq.~(\ref{W.1}).  Unlike the $M$-parity violating terms which mix the neutrinos with the neutralinos, this mixing term does not generate neutrino masses and can therefore be large in comparison (although proton decay would then dictate smaller baryon number violating interactions for the exotic triplets).  Such a large coupling would make this the most important source of $M$-parity violation thereby possibly inducing new lepton and baryon or baryon number violating decays for the LSP.

\subsection{6. Colored Triplet}
As discussed earlier, a pair of colored triplets, $\hat{T} \sim (\bar{3},1,1/3,2/3,z_T)$ and $\hat{\bar{T}} \sim (3,1,-1/3,-2/3,-6-z_T)$, are necessary for $U(1)_S$ anomaly cancellation.  The triplinos acquire mass as do the Higgsinos, from the VEV of $S$:
\begin{equation}
M_{\tilde{T}}=M_{\tilde{\bar T}}= \lambda_1 \frac{v_S}{\sqrt{2}}.
\end{equation}
The triplets themselves also accrue mass from the soft terms:
\begin{equation}
{\cal L}_{soft} \supset -m_T^2 |T|^2 \ - \ m_{\bar{T}}^2 |\bar{T}|^2 \ + \left( \ B_T T \bar{T} \ + \ \rm{h.c.} \right), 
\end{equation}
where $B_T$ is the product of a trilinear $a$-term and the VEV of $S$, and the $D$-terms.  Their physical masses are
\begin{equation}
	M_{T_{1,2}}^2 = \frac{1}{2}  \left(  \left(  M_T^2 + M_{\bar{T}}^2 \right) \mp  \sqrt{ \left(  M_T^2 - M_{\bar{T}}^2\right)^2 + 4 |B_T|^2  }   \right)  
\end{equation}
where
\begin{eqnarray}
M_T^2 &=& |\lambda_1|^2 \frac{v_S^2}{2} \ + \ m_{T}^2 \ + \ \frac{g_{BL}^2}{12} v_R^2 -  \frac{3}{8} g_S^2 v_S^2, \\
M_{\bar{T}}^2 &=& |\lambda_1|^2 \frac{v_S^2}{2} \ + \ m_{\bar{T}}^2 \ - \ \frac{g_{BL}^2}{12} v_R^2 \ + \  \frac{1}{4} g_S^2 v_S^2.
\end{eqnarray}
and we neglect electroweak $D$-term contributions.
Using the couplings of the colored triplet fields with matter allows us to write their interactions with the physical fermions. For the triplet $T$:
\begin{eqnarray}
&& \lambda_2 \ U_{3i} \ E_{3j} \ u_i \ e_j \ T, \\
&& \lambda_2 \ D_{3i} \ N_{3j} \ d_i \  \nu_j \ T, \\
&& \lambda_3 \ U^c_{3i} \ D^c_{3j} \ u_i^c \  d_j^c \ T. 
\end{eqnarray}
Here we use the standard convention for the diagonalisation of the fermion mass matrices, $U^T Y_u U^c = Y_u^{diag}$, $D^T Y_d D^c = Y_d^{diag}$, 
$E^T Y_e E^c = Y_e^{diag}$, $N^T Y_\nu N = Y_\nu^{diag}$. We also define $V=U^\dagger D$ and $V_{PMNS}=E^{\dagger} N$.  
In the case of the field $\bar{T}$ we find the following interactions:
\begin{eqnarray}
&& \lambda_4 \  U^c_{3i} \  E^c_{3j} \  u_i^c \ e^c_j  \ \bar{T}, \\
&&  \lambda_5 \ D^c_{3i} \ N^c_{3j} \ d^c_i \ \nu^c_j \ \bar{T}, \\
&& 2 \lambda_6 \ U_{3i} \ D_{3j} \  u_i d_j \ \bar{T}.
\end{eqnarray}
We are now ready to study the proton decay aspect of this theory in the next section.
\vspace{-0.4cm}
\section{VI. Phenomenological Aspects}
\subsection{A. Proton Stability}
In the previous section we have discussed the main properties of the interactions of the colored fields, $T$ and $\bar{T}$.  
Integrating out the Higgs $T$ and using the above interactions we find that the amplitude for $p \to \pi^0 e^+_\alpha$ is given 
by
\begin{equation}
{\cal A}_T (p \to \pi^0 e^+_{\alpha}) \sim \frac{\lambda_2 \lambda_3}{M_{T}^2} U^c_{31} \ D^c_{31} \ U_{31} \ E_{3 \alpha} \ < \ 10^{-30} \  \rm{GeV}^{-2}.
\end{equation}
Assuming $M_T = 1$ TeV, $U_{31}^c=U_{31}=D^c_{31}=E_{3 \alpha} \approx V_{CKM}^{13}$, one gets the bound $\lambda_3 \lambda_2 < 10^{-12}$. 
One can do something similar using the bound on the decay $p \to K^{+} \bar{\nu}_i$:
\begin{eqnarray}
{\cal A}_T (p \to K^{+} \bar{\nu}_i) &\sim& \frac{\lambda_2 \lambda_3}{M_{{T}}^2} U^c_{31} \ \left( D^c_{31} \ D_{32} +  D_{32}^c D_{31} \right) N_{3 i} \nonumber \\
 &< & \ 10^{-30} \  \rm{GeV}^{-2}.
\end{eqnarray}
The field $\bar{T}$ can mediate proton decay as well. For the channels $p \to \pi^0 e^{+}_{\alpha}$ the amplitude reads as
\begin{equation}
{\cal A}_{\bar T} (p \to \pi^0 e^+_{\alpha}) \sim \frac{2 \lambda_4 \lambda_6}{M_{\bar T}^2} U^c_{31} \ D_{31} \ U_{31} \ E^c_{3 \alpha} \ < \ 10^{-30} \  \rm{GeV}^{-2}.
\end{equation}   
Then, one gets the bound $\lambda_4 \lambda_6 < 5 \times 10^{-13}$ if $M_{\bar T}=1$ TeV and assuming $U^c_{31}=D_{31}=U_{31}=E^c_{3 \alpha} \approx V_{CKM}^{13}$. 
The same happens to the amplitude
\begin{eqnarray}
{\cal A}_{\bar T} (p \to K^{+} \bar{\nu}_i) &\sim& \frac{2 \lambda_6 \lambda_5}{M_{{\bar T}}^2} U_{31} \left( D_{32} \ D^c_{31} + D_{31}  D_{32}^c \right) N^c_{3 i} \nonumber \\
& < & 10^{-30} \  \rm{GeV}^{-2}.
\end{eqnarray}
Notice that it is difficult to set the bounds on the couplings, $\lambda_2, ..., \lambda_6$ depend on the size of the elements of flavor matrices for all quarks and leptons. 
Since the mixing between the third generation and the others two is very small in the down quark and charged lepton sectors, $D^c_{3i}=D_{3i}=E_{3i}=E^c_{3i} \approx \delta_{3i}$, 
and the bounds discussed above can be avoided. However, one has to investigate the bounds coming from proton decay at loop level. It is important to notice that 
$(V_{CKM}^*)_{i3}\approx U_{3i}$ and $N_{3i}\approx(V_{PMNS})_{3i}$, and one has contributions to the channel $p \to \pi^+ \bar{\nu}$ at two loop level:
\begin{equation}
{\cal A} (p \to \pi^{+} \bar{\nu}) \sim \frac{2 \lambda_6 \lambda_2}{(16 \pi^2)^2 M_{{T}_i}^2} \left( V_{CKM}^{13}\right)^3 \ < \ 10^{-30} \  \rm{GeV}^{-2}.
\end{equation}
In the case of $p \to K^+ \bar{\nu}$ one gets
\begin{eqnarray}
{\cal A} (p \to K^{+} \bar{\nu}) &\sim& \frac{2 \lambda_6 \lambda_2}{(16 \pi^2)^2 M_{{T}_i}^2} \left( V_{CKM}^{13}\right)^2 V_{CKM}^{32}  \nonumber \\
& < & 10^{-30} \  \rm{GeV}^{-2}.
\end{eqnarray}
Now, using this equation one can set a bound to the product: $\lambda_2 \lambda_6 <  10^{-12}$. Unfortunately, the bounds on $\lambda_3, \lambda_4$ and $\lambda_5$ depend on unknown 
mixing matrices $U^c$ and $N^c$. 
%
\subsection{B. Baryon and Lepton Number Violation at the LHC}
In this model $M$-parity is spontaneously broken after symmetry breaking and one expects the typical signals for bilinear R-parity violation.
In this subsection we will focus mainly on the properties of the new exotic fields needed to define an anomaly free theory.
The high energy analogue of the proton decay mediated by the new exotic colored triplets is potential exciting signals of baryon and lepton number violation at the LHC. 
While studies have shown that typically detecting lepton number violation at LHC is manageable, in this case it will be much more challenging since the exotic triplets couple 
only to the tau, which can decay hadronically, obscuring its lepton number.  Detecting lepton number violation would then crucially depend on how well one can see the $\tau$ leptons.  
Furthermore observing baryon number violation is always tricky at the LHC due to lack of information on the initial and final states. Specifically, the baryon number of the initial state can 
have one of five values: 0 for two gluons or for $\bar{q}q$, $\pm 1/3$ for a gluon and a quark and $\pm 2/3$ for two quarks. Therefore, one must be able to observe a final state 
with a baryon number different than these: an insurmountable task when observing light jets.  Fortunately, the fact that the exotics $T$ and $\bar{T}$ couple purely to the third generation of quarks 
and leptons in the flavor basis helps here since it is possible to tag tops and bottoms. While such issues require an in-depth study, we shall proceed by simply 
elucidating the processes that may be amiable to such a study.

Production of the colored triplets in the most efficient manner proceeds through pair production via gluon fusion. 
This is of course a strong process with large cross sections, equivalent to squark pair production from gluon fusion. 
Decay proceeds through the coupling to third generation matter; the possible final states violating baryon and lepton number are
$$
	gg \ \to \ T_i \; T_i^* \to t \;  t \;  b \;  \tau \quad \quad \text{and} \quad \quad gg \ \to \ T_i \; T_i^* \to t \;  b \;  b \;  \nu,
$$
where baryon and lepton number are both violated by one unit, as expected from proton decay operators.  
The decay width of the colored triplets depend on the size of the relevant Yukawa couplings discussed in the section 
about proton decay. Then, one can have different scenarios for given values of $\lambda_2, \lambda_3, \lambda_4, \lambda_5$, and $\lambda_6$ couplings.
For example, one can have a scenario where the main decays are into quarks and charged leptons if $\lambda_2$ 
and $\lambda_5$ are suppressed.
As for detectability, in principle at least baryon number violation can be observed in this process since the final baryon 
number of $\pm 1$ is different than any of the initial state baryon number possibilities listed above. However, 
this is crucially dependent on correctly identifying that these are all like-sign quarks.  
It goes without saying that lepton number violation can only be measured in the $\tau$ channel.
A detailed analysis of the signals is beyond the scope of this article.

\section{VII. Summary and Outlook}
We have proposed a simple model where the origin of the $\mu$ term and the matter-parity violating interactions of the MSSM 
can be understood from the spontaneous breaking of two new Abelian gauge symmetries. We have found the following results:

\begin{itemize}

\item The new symmetries are $U(1)_{B-L}$ and $U(1)_S$, where the latter is relevant only to the third generation.  In order to satisfy $U(1)_S$ anomalies new exotics, the colored triplets $T$ and $\bar{T}$, are needed.

\item The local $B-L$ gauge symmetry is broken by the VEV of the ``right-handed" sneutrinos giving rise to lepton number violating $M$-parity violation and $U(1)_S$ is broken by the VEV of $S$, generating the $\mu$-term.

\item The new $Z'$ associated with $U(1)_S$ gives rise to flavor violation without conflict with experiments.

\item We have shown that it is possible to have a consistent scenario for fermion masses after symmetry breaking. In this case one has well-defined textures for charged fermion masses 
and the mixings between the third generation and the others is very small.

\item The numerical predictions for the lightest Higgs boson have been investigated up to one-loop level showing the possibility to satisfy the experimental bounds from LEP2 experiment. 
We have found that the upper bound on the lightest Higgs mass is $m_h \sim 130$ GeV if the stop masses are below 1 TeV.

\item We made a brief discussion of how one could observe lepton and baryon number violation at the LHC in agreement with the experimental bounds on proton decay.

\end{itemize}

In our opinion this framework opens up the possibility to test the origin of the MSSM interactions ($\mu$ term and lepton number violating interactions) at the LHC.
The collider signals and the predictions for fermion masses will be investigated in a future publication.
\vspace{0.5cm}
\subsection*{Acknowledgment}
{\small P. F. P. is supported in part by the US DOE under contract No. DE-FG02-95ER40896, by the Wisconsin Alumni Research Foundation and by the James Arthur Fellowship, CCPP-New York University. 
S. S. is supported in part by the US DOE under contract No. DE-FG02-95ER40896 and by the Wisconsin Alumni Research Foundation. M. G.-A. is supported by the US DOE contract DE-FG02-08ER41531 and by the Wisconsin Alumni Research Foundation.}


\end{document}